\documentclass[a4paper,11pt]{article}
\usepackage{pos}

\title{$A=2,3$ nuclear contact coefficients in the Generalized Contact Formalism}

\author*[a,b]{Eleonora Proietti}
\author[a,b]{Laura Elisa Marcucci}
\author[b]{Michele Viviani}

\affiliation[a]{Dipartimento di Fisica "E. Fermi", Università di Pisa, Largo Pontecorvo 3, 56127 Pisa, Italy}

\affiliation[b]{Istituto Nazionale di Fisica Nucleare Sezione di Pisa, Largo Pontecorvo 3, 56127 Pisa, Italy}

\emailAdd{eleonora.proietti@phd.unipi.it}
\emailAdd{laura.elisa.marcucci@unipi.it}
\emailAdd{michele.viviani@pi.infn.it}

\abstract{
This work focuses on extracting nuclear contact coefficients for \( A = 2 \), \( A = 3 \) and \( A = 4 \) nuclei within the Generalized Contact Formalism framework. We investigate the universality of these coefficients across different nuclear systems and interaction models, using both local (in \( r \)-space) and non-local (in \( k \)-space) chiral potentials. The Hyperspherical Harmonics method is employed to calculate the nuclear wave functions from which we obtain the two-body momentum distributions and the two-body density functions, which are essential for extracting the contact coefficients. The adopted method is a rigorous ab-initio approach that can be applied to virtually any potential.

We present ratios of contact coefficients across various spin and isospin channels, highlighting their independence from the used nuclear potential. This study extends previous work where only local interaction models were employed. Furthermore, we verify whether the contact coefficient ratio between different nuclei remains consistent even when non-local potentials are considered.

Future work will extend this analysis to heavier nuclei, such as \( A = 4 \) and \( A = 6 \) nuclei.
}

\FullConference{The 11th International Workshop on Chiral Dynamics (CD2024)
\\26–30 ago 2024
\\Ruhr University Bochum, Germany
}

\begin{document}
\maketitle

\section{Introduction}

The study of short-range correlations (SRCs) in nuclei is crucial for understanding nucleon behavior at high momenta and short distances, where traditional mean-field approaches, like the nuclear shell model, fail to capture the complex dynamics of nucleon-nucleon correlations. These correlations arise from nucleon pairs with high relative momentum and low center-of-mass momentum~\cite{2018, 2021}. Furthermore, SRCs play a crucial role in defining the high-momentum tails of nuclear momentum distributions~\cite{SRC}, affecting a wide range of phenomena, from the internal structure of nucleons bound in nuclei~\cite{SRC_1, SRC_2} to the macroscopic properties of neutron stars~\cite{NS_1, NS_2}.

In the generalized contact formalism (GCF), the nuclear wave function at high momenta factorizes into a strongly interacting two-body pair and a weakly interacting residual nucleus~\cite{2021}. This formalism connects two-body momentum distributions (2BMDs) and density functions (2BDFs) to the correlated nucleon pair, reducing the contribution of the residual system to the nuclear contact coefficient. These coefficients quantify the probability of finding nucleon pairs in specific spin-isospin states and serve as a measure of SRCs across different nuclei.

Previous studies suggest that ratios of contact coefficients between different nuclei are largely independent of the employed nuclear interaction~\cite{2021}. 

While these investigations have primarily focused on local potentials~\cite{2021}, the inclusion of non-local chiral interactions, defined in \( k \)-space, is essential for a more comprehensive understanding of SRCs.

Therefore in this work, we extract nuclear contact coefficients for \( A = 2 \) and \( A = 3 \) nuclei within the GCF framework, employing a broad range of chiral interactions, both local (in \( r \)-space) and non-local (in \( k \)-space). Additionally, we use the Hyperspherical Harmonics (HH) method, a rigorous ab-initio approach capable of computing nuclear wave functions and consequently 2BMD and 2BDF with virtually any potential~\cite{HH, HH-2008}. By analyzing ratios of contact coefficients across various spin and isospin channels, we aim to evaluate the universality of SRCs in nuclei.

\section{Theoretical Formalism}
\subsection{Two-body Momentum Distributions and Density Functions}

The probability of finding two nucleons, $N_1$ and $N_2$, with relative momentum $k$ in a given nucleus is proportional to the 2BMD~\cite{2bmd}, expressed as
\begin{align}
    n_{N_1N_2}({k}) =  \int d\hat{\textbf{k}} \int d{\textbf{K}} \; \psi^\dagger(\textbf{k},\textbf{K}) \, P_{N_1 N_2} \, \psi(\textbf{k},\textbf{K}), \label{2bmd_NN}
\end{align}
where $\psi(\textbf{k},\textbf{K})$ represents the nuclear wave function obtained using the HH method~\cite{HH, HH-2008}, $P_{N_1 N_2}$ is the projector onto the nucleon pair $N_1 N_2 = pn, pp, nn$, $\textbf{k}$ is the relative momentum while $\textbf{K}$ is the pair center of mass momentum. The dependence and the integration over the coordinates of
particles 3-A is understood.

\medskip

It is also possible to define the probability of finding two nucleons with relative momentum $k$ and spin-isospin state $ST$ using the spin-isospin projector $P^{ST}$ as
\begin{equation}
    n^{ST}({k}) = \int d \hat{\textbf{k}} \int d \textbf{K} \; \psi^{\dagger}(\textbf{k},\textbf{K}) P^{ST}\psi(\textbf{k},\textbf{K}). \label{2bmd_ST}
\end{equation}
Similarly, the 2BMD for two nucleons \( N_1 \) \( N_2 \) with relative momentum \( k \) in a given spin state \( S \) can be defined as
\begin{align}
        n_{N_1\, N_2}^S({k}) = \int d \hat{\textbf{k}} \int d \textbf{K} \; \psi^{\dagger}(\textbf{k},\textbf{K})  P_{N_1\, N_2}^S \psi(\textbf{k},\textbf{K}) \label{2bmd_S}.
\end{align}
Working in coordinate space, analogous definitions can be formulated for the 2BDFs.
For instance, the 2BMD for two nucleons $N_1 \, N_2$ with relative distance $r$ in a given spin state $S$ is defined as
\begin{align}
        \rho_{N_1\, N_2}^S({r}) = \int d \hat{\textbf{r}} \int d \textbf{R} \; \psi^{\dagger}(\textbf{r},\textbf{R})  P_{N_1\, N_2}^S \psi(\textbf{r},\textbf{R}) \label{2bdf_S},
\end{align}
where $P_{N_1\, N_2}^S$ is the same projection operator of Eq.~\eqref{2bmd_S}, $\textbf{r}$ is the relative position and $\textbf{R}$ is the pair center of mass position.

\subsection{Interaction Models}\label{sez}

The $A=3$ wave functions $\psi(k,K)$ or $\psi(r,R)$ have been calculated using the HH method. Due to lack of space, we refer to the reviews of Refs.~\cite{HH, HH-2008}. Here we only remark that the HH method can be used with essentially any type of two-nucleon (NN) potential, both local and non-local. At present, though, the method can be used with only local three-nucleon (3N) interactions. The variety of interaction models adopted in this study is presented below. 

The first interaction model used is the AV18 NN interaction~\cite{av18} complemented by the Urbana IX 3N force~\cite{uix} (AV18/UIX). So that our results can be validated against the ones in Ref.~\cite{2021}. Additionally, we have employed various chiral potentials.

In particular, we have used the local Norfolk chiral potentials, including NN and 3N forces, labeled NV2+3/Ia*, NV2+3/IIa*, NV2+3/Ib*, NV2+3/IIb* as derived in Refs.~\cite{Norfolk-2,Norfolk-3}. 

Furthermore, we have considered the non-local chiral NN interactions of Ref.~\cite{EMN}, derived at the chiral order N2LO, N3LO, and N4LO, with different cutoffs (\( \Lambda = 450, 500, 550 \, \text{MeV} \)). To these NN interaction models, we have added the local 3N interaction at N2LO with the $c_D-c_E$ low energy constants derived in Ref~\cite{LEC}.  
All the adopted models are able to reproduce the $A=2,3,4$ binding energies, as shown in Refs.~\cite{HH, HH-2008}, applying the HH method. However, at present, we are unable to verify whether the order mismatch between the NN and 3N interaction for non-local cases, as well as the differences in the cutoff, which lead to non-local NN and local 3N interactions, might affect our results.



\subsubsection{Nucleus-dependent nuclear contact coefficients}

In the 2BMDs (2BDFs), SRCs are dominant at large \( k \) (small \( r \)). These regions are expected to be well described by the GCF~\cite{2018}. Within the GCF framework, the 2BDFs (or 2BMDs) are expressed in terms of the so-called universal function (or its Fourier Transform, FT) and the nuclear contact coefficients, acceding to the expression
\begin{align}
    n_{N_1 N_2, \; A}^S(k) & \xrightarrow{k \to \infty} \Tilde{C}_{N_1 N_2, \; A}^S \, |\Tilde{\varphi}^S_{N_1 N_2}(k)|^2, \label{2bmd} \\
    \rho_{N_1 N_2, \; A}^S(r) & \xrightarrow{r \to 0} {C}_{N_1 N_2, \; A}^S \, |{\varphi}^S_{N_1 N_2}(r)|^2. \label{2bdf}
\end{align}
Here, the universal function ${\varphi}^S_{N_1 N_2}(r)$ (or its FT $\Tilde{\varphi}^S_{N_1 N_2}(k)$) represents the correlated nucleon pair and depends solely on the interaction and the spin channel, while the nuclear contact coefficients ${C}_{N_1 N_2, \; A}^S$ and $\Tilde{C}_{N_1 N_2, \; A}^S$ account for the rest of the system, depending on both the nucleus and the interaction.
More specifically, in the \( S = 0 \) channels, the universal function corresponds to the zero-energy two-body scattering states with \( L = S = J = 0 \). For \( S = 1 \), it corresponds to the deuteron ground state (\( L = 0, 2 \), \( S = 1 \), \( J = 1 \)). Note that we have added the subscript $A$, to indicate that Eqs.~\eqref{2bmd} and~\eqref{2bdf} can be applied to various nuclei. 

\subsubsection{Extraction of the nuclear contact coefficient}

In this work, we focus on the contact coefficients \( C_{nn/pp, \; A}^{S=0} \), \( C_{np, \; A}^{S=1} \), and \( C_{np, \; A}^{S=0} \) for \( A = 2, 3 \). Furthermore, we assume the Coulomb interaction to be negligible, so that \( C_{nn, \; A}^{S=0} \equiv C_{pp, \; A}^{S=0} \).

Initially, the contact coefficients have been extracted for each nucleus \( A \) following the procedure of Ref.~\cite{2021}, i.e. assuming the relative angular momentum $\ell$ to be zero, so that $S=0$ corresponds to the $ST=0,1$ state, and $S=1$ corresponds to the $ST=1,0$ state. Consequently, the contact coefficients are defined as
\begin{align}
    \Tilde{C}_{pp/nn}^{S=0}(k) & = \lim\limits_{k \to \infty} \, \left\{n_{pp/nn}({k}) \; / \; |\Tilde{\varphi}_{pp/nn}^{S=0}({k})|^2\right\}, \label{Cpp0_old_k} \\
    \Tilde{C}^{S=1}_{pn}(k) & = \lim\limits_{k \to \infty} \, \left\{ n^{ST=10}({k}) \; / \; |\Tilde{\varphi}_{pn}^{S=1}({k})|^2\right\}, \label{Cnp0_old_k} \\
    \Tilde{C}^{S=0}_{pn}(k) & = \lim\limits_{k \to \infty} \, \left\{\left[{n_{pn}({k}) - \Tilde{C}^{S=1}_{pn} |\Tilde{\varphi}_{pn}^{S=1}({k})|^2}\right] / |\Tilde{\varphi}_{pn}^{S=0}({k})|^2\right\}. \label{Cnp1_old_k}
\end{align}
when using the 2BMDs, and
\begin{align}
    C_{pp/nn}^{S=0}(r) & = \lim\limits_{r \to 0} \, \left\{\rho_{pp/nn}({r}) \; / \; |\varphi_{pp/nn}^{S=0}({r})|^2\right\}, \label{Cpp0_old_r} \\
    C^{S=1}_{pn}(r) & = \lim\limits_{r \to 0} \, \left\{\rho^{S=1}_{pn}({r}) \; / \; |\varphi_{pn}^{S=1}({r})|^2\right\}, \label{Cnp0_old_r} \\
    C^{S=0}_{pn}(r) & = \lim\limits_{r \to 0}\, \left\{\left[{\rho_{pn}({r}) - C^{S=1}_{pn} |\varphi_{pn}^{S=1}({r})|^2}\right] / |\varphi_{pn}^{S=0}({r})|^2\right\}. \label{Cnp1_old_r}
\end{align}
for the 2BDFs.
A more general definition, which goes beyond the assumption \( \ell = 0 \), is obtained calculating the 2BMDs and 2BDFs of two nucleons \( N_1 \) \( N_2 \) in a given spin state \( S \). Therefore we can write
\begin{align}
    \Tilde{C}_{pp/nn}^{S=0}(k) &  = \lim\limits_{k \to \infty} \, \left\{ n_{pp/nn}^{S=0}({k}) \; / \; |\Tilde{\varphi}_{pp/nn}^{S=0}({k})|^2 \right\}, \label{Cpp0_k} \\
    \Tilde{C}^{S=1}_{pn}(k) &  = \lim\limits_{k \to \infty} \, \left\{ n^{S=1}_{pn}({k}) \; / \; |\Tilde{\varphi}_{pn}^{S=1}({k})|^2 \right\}, \label{Cnp0_k} \\
    \Tilde{C}^{S=0}_{pn}(k) &  = \lim\limits_{k \to \infty} \, \left\{ n^{S=0}_{pn}({k}) / |\Tilde{\varphi}_{pn}^{S=0}({k})|^2 \right\}. \label{Cnp1_k}
\end{align}
and similarly
\begin{align}
    C_{pp/nn}^{S=0}(r) & = \lim\limits_{r \to 0} \, \left\{\rho_{pp/nn}^{S=0}({r}) \; / \; |\varphi_{pp/nn}^{S=0}({r})|^2 \right\}, \label{Cpp0_r} \\
    C^{S=1}_{pn}(r) & = \lim\limits_{r \to 0} \, \left\{ \rho^{S=1}_{pn}({r}) \; / \; |\varphi_{pn}^{S=1}({r})|^2 \right\}, \label{Cnp0_r} \\
    C^{S=0}_{pn}(r) & = \lim\limits_{r \to 0} \, \left\{ \rho^{S=0}_{pn}({r}) / |\varphi_{pn}^{S=0}({r})|^2 \right\}. \label{Cnp1_r}
\end{align}

The contact coefficients are extracted from the ratio \( n(k) / |\Tilde{\varphi}(k)|^2 \) for large values of \( k \), and \( \rho(r) / |\varphi(r)|^2 \) for small values of \( r \), and we expected in these regions to have a plateau in the ratios \( n(k) / |\Tilde{\varphi}(k)|^2 \) and \( \rho(r) / |\varphi(r)|^2 \).

\subsection{Ratios of nuclear contact coefficients}

According to the GCF~\cite{2021}, the contact coefficients ${C}_{N_1 N_2, \; A}^S$ and $\Tilde{C}_{N_1 N_2, \; A}^S$ should be the same, i.e. the contact coefficients should be independent on whether we work in $k-$ or in $r-$space. Furthermore, the ratios of contact coefficients with respect to a reference nucleus $A_0$ should be model-independent~\cite{2021}. Here we study the ratios
\begin{align*}
    & \; {C}_{N_1 N_2, \; A}^S /{C}_{N_1 N_2, \; A_0}^S, \\ 
   & \; \Tilde{C}_{N_1 N_2, \; A}^S/\Tilde{C}_{N_1 N_2, \; A_0}^S,
\end{align*}
for \( S = 1 \), where \( A_0 \) is chosen to be the deuteron (\( d \)). Note that the \( S = 0 \) case, for which \( A_0 \) is the \( ^4\text{He} \) nucleus, is still under investigation.

As shown by Cruz-Torres et al. in Ref.~\cite{2021}, the ratios for a given nucleus remain consistent within uncertainties across different potentials.

Since the ratio is independent of the interaction, "soft" potentials can be used for heavy nuclei. Specifically, \( {C}_{N_1 N_2, \; A_0}^S \) is calculated with a "hard" potential, and the contact coefficient for heavier nuclei is inferred by multiplying the "soft" potential ratio \( {C}_{N_1 N_2, \; A}^S / {C}_{N_1 N_2, \; A_0}^S \) with \( {C}_{N_1 N_2, \; A_0}^S \) calculated using a 'hard' potential.

The results in Ref.~\cite{2021} suggest that the ratios are independent of the specific interaction used but depend on the considered nucleus. The ratios of $r-$space or $k-$space coefficients are also consistent. These results were obtained by using only local potentials. Our results shown in the next section confirm, at least for the cases studied here, that this behavior holds for any potential, both local and non-local.

\section{Results}

\begin{figure}[htbp]
\centering
\includegraphics[width=1.35\linewidth]{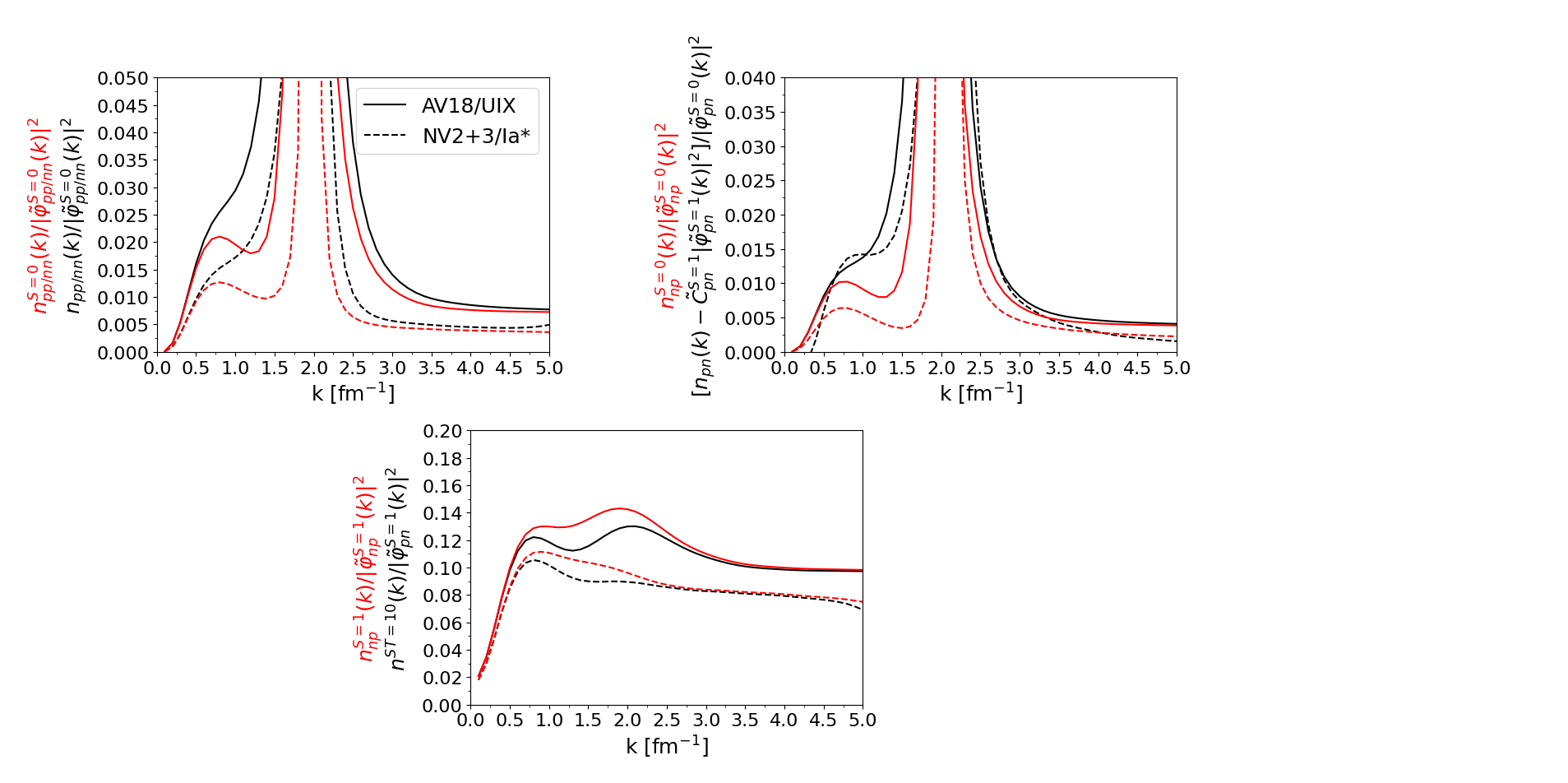}
\caption{Ratios of the 2BMDs and the FT of the universal function, as defined in Eqs.~\eqref{Cpp0_old_k}--~\eqref{Cnp1_old_k} (with the $\ell=0$ assumption) and~\eqref{Cpp0_k}--~\eqref{Cnp1_k} (without the $\ell=0$ assumption), for the local potentials AV18/UIX and NV2+3/Ia*. The black curve represents calculations with \( \ell = 0 \), while the red curve shows results without this assumption.}
\label{fig:fig1}
\end{figure}

\begin{figure}[htbp]
\centering
\includegraphics[width=1.35\linewidth]{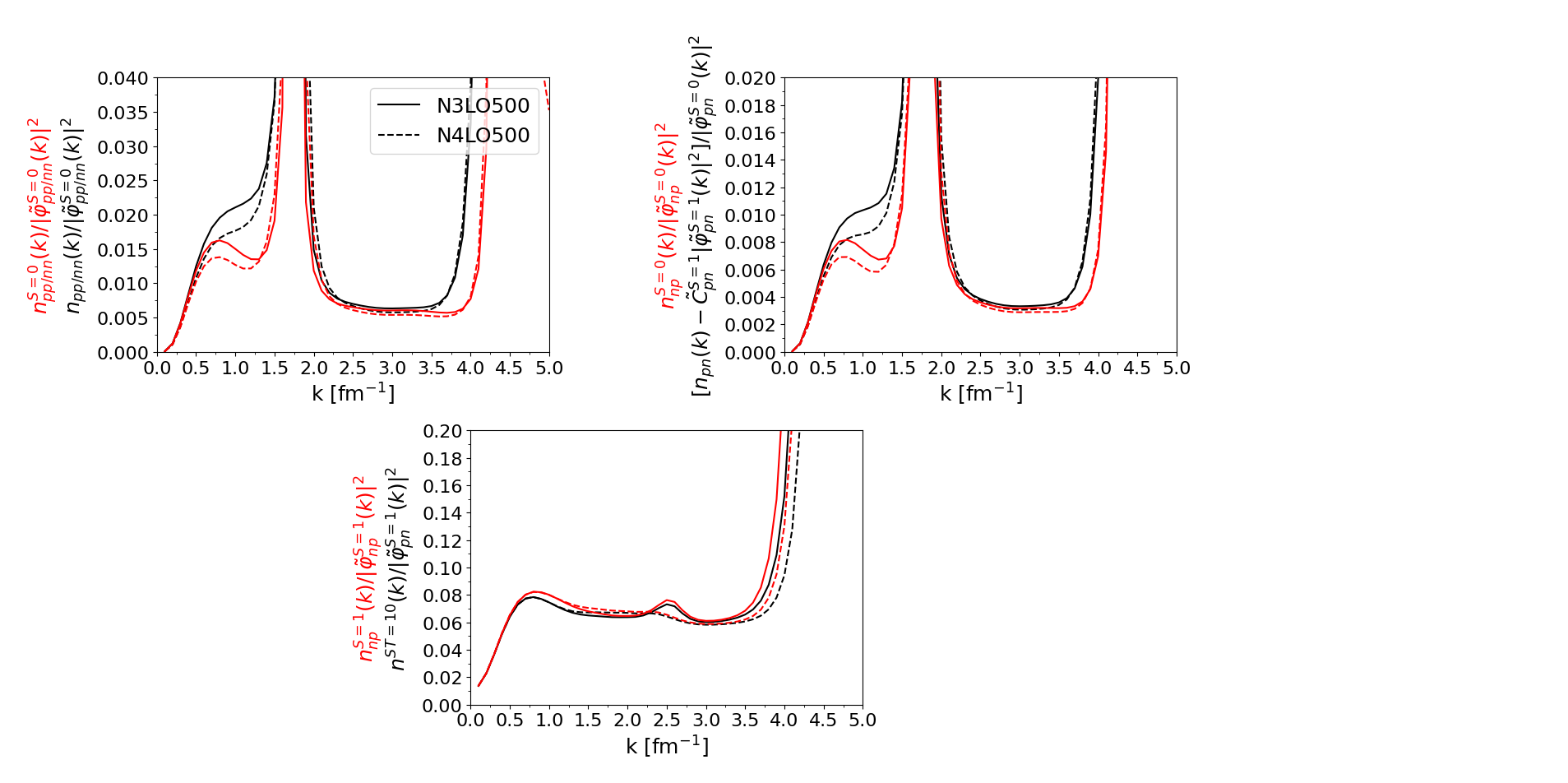}
\caption{Same as Fig.~\ref{fig:fig1}, but for the non-local potentials N3LO500 and N4LO500.}
\label{fig:fig2}
\end{figure}

\begin{figure}[htbp]
\centering
\includegraphics[width=1.35\linewidth]{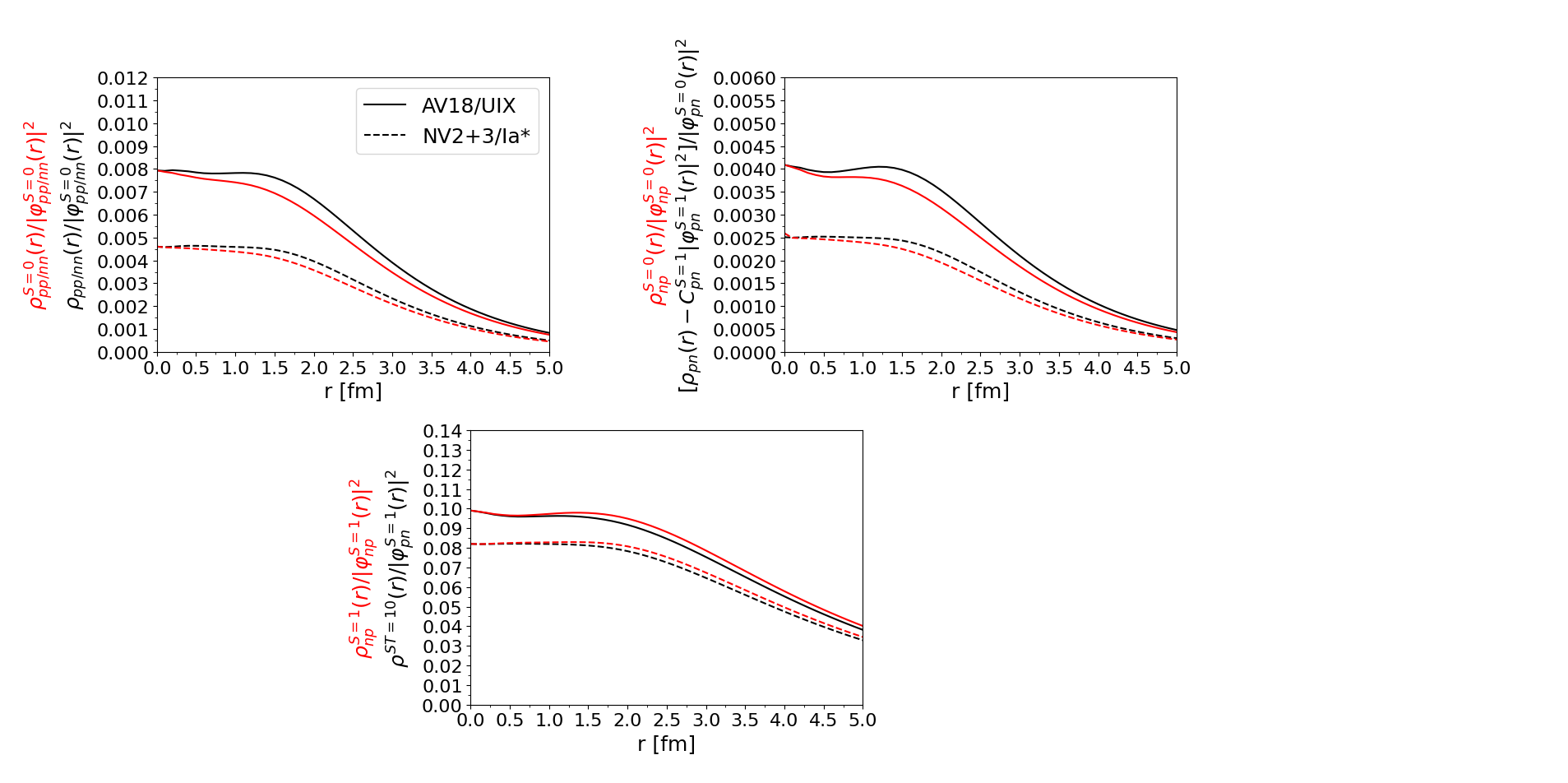}
\caption{Same as Fig.~\ref{fig:fig1} but for the ratios of the 2BDFs and the universal functions, as defined in Eqs.~\eqref{Cpp0_old_r}--~\eqref{Cnp1_old_r} (with the $\ell=0$ assumption) and~\eqref{Cpp0_r}--~\eqref{Cnp1_r} (without the $\ell=0$ assumption).}
\label{fig:fig3}
\end{figure}

\begin{figure}[htbp]
\centering
\includegraphics[width=1.35\linewidth]{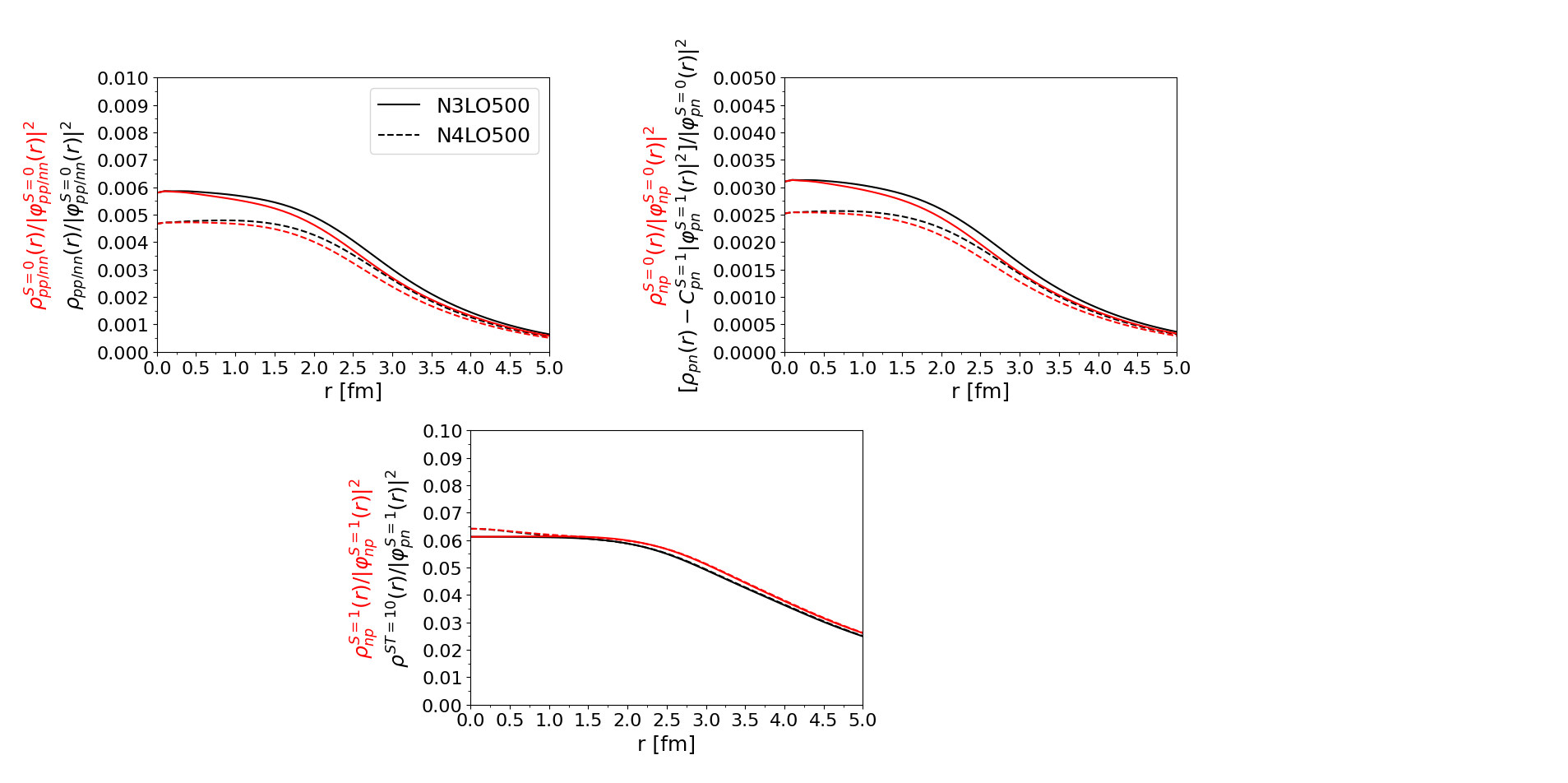}
\caption{Same as Fig.~\ref{fig:fig3}, but for the non-local potentials N3LO500 and N4LO500.}
\label{fig:fig4}
\end{figure}

As a first step, in Figs.~\ref{fig:fig1}--\ref{fig:fig4} we present the ratios of the 2BMDs to the FT of the universal function in \( k \)-space, as defined in Eqs.~\eqref{Cpp0_old_k}--~\eqref{Cnp1_old_k} or, without the assumption $\ell=0$, in Eqs.~\eqref{Cpp0_k}--~\eqref{Cnp1_k}, as well as the ratios of the 2BDFs to the universal function in \( r \)-space, as defined in Eqs.~\eqref{Cpp0_old_r}--~\eqref{Cnp1_old_r} or in Eqs.~\eqref{Cpp0_r}--~\eqref{Cnp1_r} without the $\ell=0$ assumption. The considered channels are \( nn \) with \( S=0 \), \( np \) with \( S=0 \), and \( np \) with \( S=1 \). These ratios have been calculated for all local and non-local potentials listed in Sez.~\ref{sez}, but in the figures we give representative examples, i.e. the AV18/UIX and NV2+3/Ia* for local potentials, and the N3LO500 and N4LO500 for non-local potentials.
As expected, in Figs.~\ref{fig:fig1}--\ref{fig:fig4}, plateaus are observed for large values of \( k \) and small values of \( r \), corresponding to regions dominated by SRCs. Specifically, in \( k \)-space, the plateau region for local potentials occurs at very high \( k \) (beyond 3.5--4 fm\(^{-1}\)). For non-local potentials, the plateau region is found for slightly smaller values of \( k \).
The divergences at high momenta are cutoff effects, and the corresponding range of $k$ cannot be considered. In \( r \)-space, the plateau region for both local and non-local potentials appears consistently at \( r \to 0 \).

Inspecting Figs.~\ref{fig:fig1}--\ref{fig:fig4}, we can also observe that the differences between the calculations with the \( \ell=0 \) approximation (black curves) and without it (red curves) are evident. However, in the plateau region, these differences are small, confirming that \( s \)-waves give the dominant contribution to SRCs, as expected. However, noticeable variations are still observed, particularly in \( k \)-space, where the plateaus without the \( \ell=0 \) approximation appear for a wider range of values of $k$, underscoring the importance of going beyond this assumption.
Moreover, by comparing the contact coefficients extracted in the plateau region, which are not reported in this work due to space limitations, it can be easily observed that the dominant channel is the \( np \, S=1 \) channel, with the SRCs predominantly located in the so-called deuteron channel.

\begin{figure}
    \centering
    \includegraphics[width=0.7\linewidth]{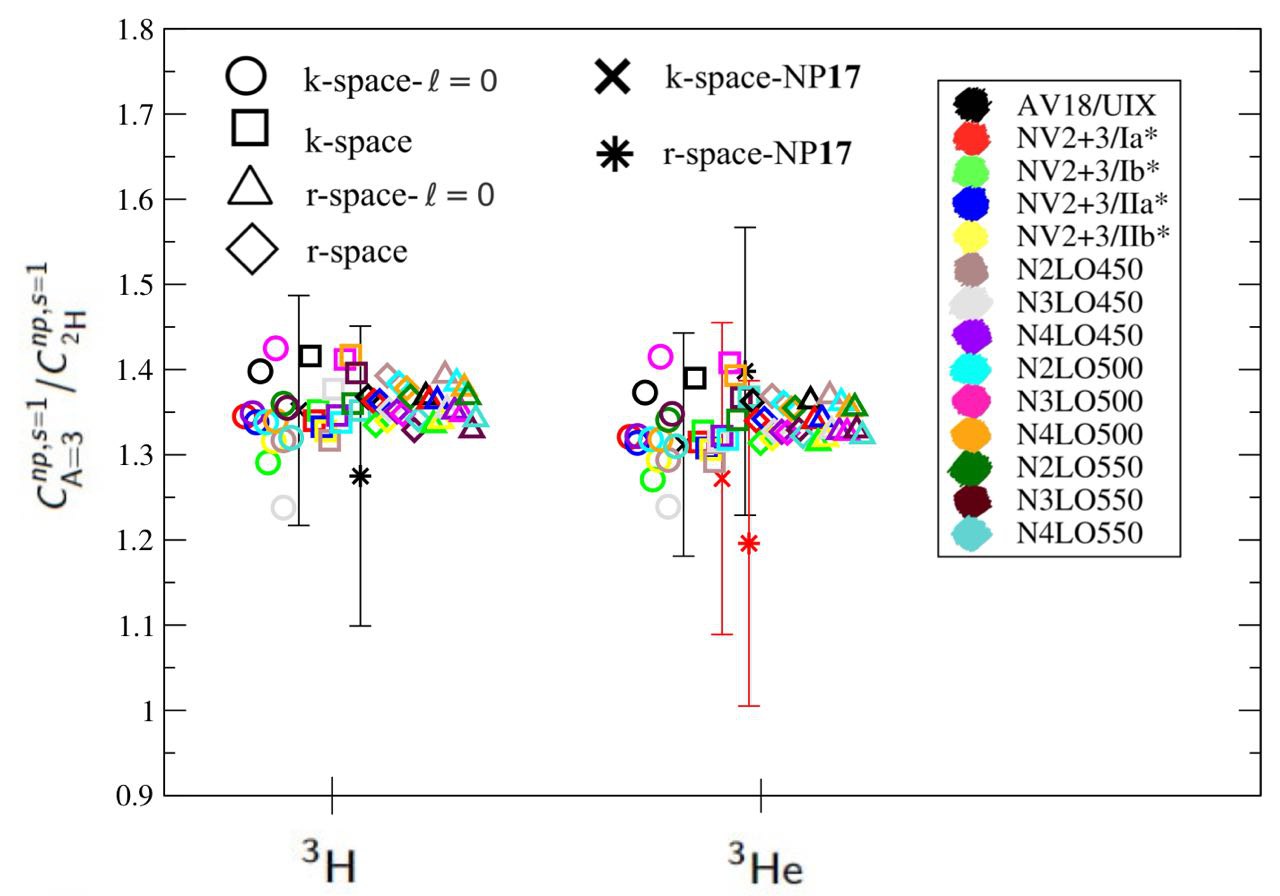}
 \caption{Ratios of the contact coefficients 
\(\Tilde{C}_{np, A=3}^{S=1}/\Tilde{C}_{np, \text{d}}^{S=1}\) 
or \({C}_{np, A=3}^{S=1}/{C}_{np, \text{d}}^{S=1}\) for the \(^{3}\text{H}\) and \(^{3}\text{He}\) nuclei, with all the nuclear interaction models of Sec.~\ref{sez}. The points labeled NP\textbf{17} correspond to the results from Ref.~\cite{2021}.}
    \label{fig:ratio}
\end{figure}

In Fig.~\ref{fig:ratio}, we show the ratios of contact coefficients for the dominant \( np \, S=1 \) channel (\({C_{np, \; A=3}^{S=1}}/{C_{np, \; \text{d}}^{S=1}}\)), including calculations with and without the \( \ell=0 \) approximation for all previously described potentials (see Sec.~\ref{sez}). The consistency of the results across different calculations and approximations emphasizes their independence from the choice of using 2BMDs or 2BDFs. The results exhibit minimal sensitivity to the specific potential employed, whether local or non-local, further supporting the hypothesis of model-independence of the ratios across different interaction models. Additionally, our results are in good agreement with those of Ref.~\cite{2021}.


\section{Conclusions and outlook}

In this work, we have applied the GCF and extracted the nuclear contact coefficients from the 2BMDs and 2BDFs, using a variety of interaction models, both local and non-local. We have focused on the $A=2,3$ nuclei. The results highlight the robustness of the GCF, demonstrating that the ratios \({C_{np, \; A=3}^{S=1}}/{C_{np, \; \text{d}}^{S=1}}\) are largely independent of the specific nuclear interaction employed~\cite{2018,2021}. This universality suggests that the nuclear contact coefficients are predominantly determined by short-range correlations between nucleons, rather than by the details of the interaction models.



Several developments are planned. In the short term, we aim to extend our analysis to $^4$He~\cite{HH,HH-2008}, in order to compute the ratios of contact coefficients for the \(S=0\) channels. 
 Furthermore, we plan to establish a robust procedure that will allow us to estimate the errors associated with the extraction of these coefficients. This will allow us to access the model-independence as well as the $r$- or $k$-space equivalence of our results. 

In the long term, we aim to extend our study to the \(A = 6\) nuclei using the HH method~\cite{A=6}. This will provide a deeper understanding of the behavior of short-range correlations in heavier nuclei, and further extend the applicability of the GCF to more complex nuclear systems.



\begin{thebibliography}{99}

\bibitem{2018}{ R. Weiss, et al., \emph{“The nuclear contacts and short-range correlations in nuclei”}, Phys. Lett. B \textbf{780}, 211 (2018).}

\bibitem{2021}{R. Cruz-Torres et al., \emph{“Many-body factorization and position–momentum equivalence of nuclear short-range correlations”}, Nat. Phys. \textbf{17}, 306 (2021).} 

\bibitem{SRC}{O. Hen et al.,
 \emph{"Nucleon-nucleon correlations, short-lived excitations, and the quarks within"}, Rev. Mod. Phys. \textbf{89}, 045002 (2017). }

 \bibitem{SRC_1}{B. Schmookler et al.,
\emph{"Modified structure of protons and neutrons in correlated pairs"}, Nat. Phys. \textbf{566}, 354 (2019).}

\bibitem{SRC_2}{E. P. Segarra et al.,
\emph{"Neutron valence structure from nuclear deep inelastic scattering"}, Phys. Rev. Lett. \textbf{124}, 092002 (2020).}

\bibitem{NS_1}{L. Frankfurt, M. Sargsian, and M. Strikman,
   \emph{"Recent observation of short range nucleon correlations in nuclei and their implications for the structure of nuclei and neutron stars"}, Int. J. Mod. Phys. A \textbf{23}, 2991 (2008).}

\bibitem{NS_2}{B.-A. Li et al.,
\emph{"Nucleon Effective Masses in Neutron-Rich Matter"}, Prog. Part. Nucl. Phys. \textbf{99}, 29 (2018) 29.}

\bibitem{HH-2008}{A. Kievsky et al.,
\emph{“A High-precision variational approach to three- and four-nucleon bound and zero-energy scattering states”}, J. Phys. G \textbf{35}, 063101 (2008).}

\bibitem{HH}{ L. E. Marcucci et al., 
\emph{“The Hyperspherical Harmonics Method: A Tool for Testing and Improving Nuclear Interaction Models”}, Front. in Phys \textbf{8}, 69 (2020).}

\bibitem{2bmd}{ L. E. Marcucci et al., 
\emph{“Momentum distributions and short-range correlations in the deuteron and $^3He$ with modern chiral potentials”}, Phys. Rev. C \textbf{99}, 034003 (2019).}

\bibitem{av18}{R. B. Wiringa, V. G. J. Stoks, and R. Schiavilla, 
\emph{“An Accurate nucleon-nucleon potential with charge independence breaking”}, Phys. Rev. C \textbf{51}, 38 (1995).}

\bibitem{uix}{B. S. Pudliner et al., 
\emph{“Quantum Monte Carlo Calculations of A$\le$6 Nuclei”}, Phys. Rev. Lett. \textbf{74}, 4396 (1995).}

\bibitem{Norfolk-2}{M. Piarulli et al.,
\emph{“Minimally non local nucleon-nucleon potentials with chiral two-pion exchange including $\Delta$ resonances”}, Phys. Rev. C \textbf{91}, 024003 (2015).}

\bibitem{Norfolk-3}{A. Baroni et al.,
\emph{“Local chiral interactions, the tritium Gamow-Teller matrix element, and the three-nucleon contact term”}, Phys. Rev. C \textbf{98}, 044003 (2018).}

\bibitem{LEC}{A. Gnech, L. E. Marcucci, and M. Viviani,
\emph{“Bayesian analysis of muon capture on the deuteron in chiral effective field theory”}, Phys. Rev. C \textbf{109}, 035502 (2024).}

\bibitem{EMN}{D. R. Entem, R. Machleidt, and Y. Nosyk,
\emph{“High-quality two-nucleon potentials up to fifth order of the chiral expansion”}, Phys. Rev. C \textbf{96}, 024004 (2017).}

\bibitem{A=6}{A. Gnech, M. Viviani, and L. E. Marcucci,
\emph{"Calculation of the $^6$Li ground state within the hyperspherical harmonic basis"}, Phys. Rev. C \textbf{102}, 014001 (2020).}

\end{thebibliography}
\end{document}